\documentclass[10pt,journal, twocolumn]{IEEEtran}
\usepackage{url}
\usepackage{graphicx}
\usepackage{epstopdf}
\usepackage{algorithm}
\usepackage{algorithmicx}
\usepackage{algpseudocode}
\usepackage{cite}
\usepackage{color}
\usepackage{amsmath,amssymb,amsfonts,amsthm}
\usepackage{bm}
\usepackage{setspace}
\usepackage{multirow}
\usepackage{mathtools}
\usepackage[table,xcdraw]{xcolor}
\usepackage{capt-of}

\newcommand{\asconv}{\overset{\text{a.s.}}{\longrightarrow}}
\newcommand{\ra}{\rightarrow}
\newcommand{\LB}{\left\{}
\newcommand{\RB}{\right\}}
\newcommand{\Lb}{\left[}
\newcommand{\Rb}{\right]}
\newcommand{\lb}{\left(}
\newcommand{\rb}{\right)}

\newcommand{\pone}{p_1}
\newcommand{\ptwo}{p_2}
\newcommand{\none}{{n_1}}
\newcommand{\ntwo}{{n_2}}
\newcommand{\hnone}{\widehat{n}_1}
\newcommand{\hntwo}{\widehat{n}_2}

\newcommand{\ASone}{\mathbf{A}_{S_1}}
\newcommand{\AStwo}{\mathbf{A}_{S_2}}

\newcommand{\bDeltaS}{\mathbf{\Delta}_S}
\newcommand{\bDeltaN}{\mathbf{\Delta}_N}
\newcommand{\bA}{\mathbf{A}}
\newcommand{\bAN}{{\mathbf{A}_N}}
\newcommand{\bAS}{{\mathbf{A}_S}}
\newcommand{\bD}{\mathbf{D}}

\newcommand{\bL}{\mathbf{L}}
\newcommand{\bLN}{{\mathbf{L}_N}}
\newcommand{\bLS}{{\mathbf{L}_S}}

\newcommand{\bCS}{\mathbf{C}_S}
\newcommand{\bCN}{\mathbf{C}_N}

\newcommand{\bCSbar}{\mathbf{\bar{C}}_S}
\newcommand{\bCNbar}{\mathbf{\bar{C}}_N}

\newcommand{\bone}{\mathbf{1}}
\newcommand{\onenone}{\mathbf{1}_{\none}}
\newcommand{\onentwo}{\mathbf{1}_{\ntwo}}

\newcommand{\DNone}{\mathbf{D}_{N_1}}
\newcommand{\DNtwo}{\mathbf{D}_{N_2}}
\newcommand{\DSone}{\mathbf{D}_{S_1}}
\newcommand{\DStwo}{\mathbf{D}_{S_2}}
\newcommand{\Lone}{\mathbf{L}_1}
\newcommand{\Ltwo}{\mathbf{L}_2}
\newcommand{\LNone}{\mathbf{L}_{N_1}}
\newcommand{\LNtwo}{\mathbf{L}_{N_2}}
\newcommand{\LSone}{\mathbf{L}_{S_1}}
\newcommand{\LStwo}{\mathbf{L}_{S_2}}
\newcommand{\yone}{\mathbf{y}_1}
\newcommand{\ytwo}{\mathbf{y}_2}
\newcommand{\xone}{\mathbf{x}_1}
\newcommand{\xtwo}{\mathbf{x}_2}
\newcommand{\bx}{\mathbf{x}}
\newcommand{\by}{\mathbf{y}}
\newcommand{\pUB}{p_{\text{UB}}}
\newcommand{\pLB}{p_{\text{LB}}}
\newcommand{\hp}{\widehat{p}}
\newcommand{\hpUB}{\widehat{p}_{\text{UB}}}
\newcommand{\hpLB}{\widehat{p}_{\text{LB}}}
\newcommand{\hLone}{\widehat{\mathbf{L}}_1}
\newcommand{\hLtwo}{\widehat{\mathbf{L}}_2}

\begin{document}
\title{Phase Transitions in Spectral Community Detection of Large Noisy Networks}

\author{Pin-Yu~Chen and Alfred O. Hero III,~\emph{Fellow},~\emph{IEEE}
\\ Department of Electrical Engineering and Computer Science, University of Michigan, Ann Arbor, USA
\\Email : \{pinyu,hero\}@umich.edu
\thanks{This work has been partially supported by the Army Research Office (ARO), grant number W911NF-12-1-0443.}
}

\pagestyle {empty}
\thispagestyle{empty}

\maketitle
\thispagestyle{empty}
\small
\begin{abstract}
In this paper, we study the sensitivity of the spectral clustering based community detection algorithm subject to a Erdos-Renyi type random noise model. We prove phase transitions in community detectability as a function of the external edge connection probability and the noisy edge presence probability under a general network model where two arbitrarily connected communities are interconnected by random external edges. Specifically, the community detection performance transitions from almost perfect detectability to low detectability as the inter-community edge connection probability exceeds some critical value. We derive upper and lower bounds on the critical value and show that the bounds are identical when the two communities have the same size. The phase transition results are validated using network simulations. Using the derived expressions for the phase transition threshold we propose a method for estimating this threshold from observed data.
\end{abstract}


\section{Introduction}
\label{sec_Intro}
Community detection is a graph signal processing problem \cite{Fortunato10,Miller10,Sandryhaila13,Bertrand13,Shuman13,MillerICASSP10,CPY14deep,SihengChen14,CPY14ICASSP} where the goal is to cluster the nodes on a graph into different communities by inspecting the connectivity structure of the graph.
Consider an undirected regular graph consisting of two node-disjoint communities interconnected by some external edges. Let $n$ denote the total number of nodes in the network. The network topology can be characterized by its symmetric adjacency matrix $\bA$, where $\bA$ is an $n \times n$ matrix, with $\bA_{ij}=1$ if an edge exists between nodes $i$ and $j$, and $\bA_{ij}=0$ otherwise.

Since community detection can be viewed as a graph partitioning problem that can be solved by identifying the graph cut that correctly separates the communities, spectral clustering \cite{Luxburg07,Shi00} approaches to community detection are natural \cite{White05,Gennip12,Tsironis13,Huang14}.
Spectral clustering specifies a graph cut by inspecting the eigenstructure of the graph. Let $\mathbf{1}_n(\mathbf{0}_n)$ be the $n$-dimensional all-one (all-zero) vector.
Define $\bL=\bD-\bA$ as the graph Laplacian matrix of the graph, where
$\bD=\text{diag}(\bA \mathbf{1}_n)$ is the diagonal degree matrix.
Let $\lambda_i(\bL)$ denote the $i$-th smallest eigenvalue of $\bL$. It is well-known that $\lambda_1(\bL)=0$ since $\bL \mathbf{1}_n=\mathbf{0}_n $ and $\bL$ is a positive semidefinite (PSD) matrix \cite{Merris94,Chung97SpectralGraph}. The second smallest eigenvalue, $\lambda_2(\bL)$, is known as the algebraic connectivity. The eigenvector associated with $\lambda_2(\bL)$ is called the Fiedler vector \cite{Fiedler73}. A mathematical representation of the algebraic connectivity is
\begin{align}
\label{eqn_alge}
\lambda_2(\bL)= \min_{\|\bx\|_2=1,\mathbf{1}_n^T \bx=0} \bx^T \bL \bx.
\end{align}

The principle underlying spectral clustering for community detection \cite{White05,Gennip12,Tsironis13,Huang14} is summarized as follows:
\begin{enumerate}
  \item Compute the graph Laplacian matrix $\mathbf{L}=\mathbf{D}-\mathbf{A}$.
  \item Compute the Fiedler vector $\by$.
  \item Perform K-means clustering \cite{Hartigan1979} on the entries of $\by$ to cluster the nodes into two groups. To detect more than two communities, we can use successive spectral clustering on the discovered communities \cite{Newman06community,Fortunato10}.
\end{enumerate}

 Most literature on community detectability \cite{Bickel09,Zhao12,Nadakuditi12Detecability,Krzakala13,Radicchi13_hetero,Radicchi14,CPY14spectral,CPY14modularity} focuses on the noiseless setting where the edges are not subject to random insertions or deletions. However, in practice the network data can be corrupted by incorrect measurements or background noises (e.g., bio-informatics data) that can produce such random insertions and deletions. Consequently, analyzing the sensitivity of community detection algorithms to noise is an important task.
In this paper, we prove the existence of abrupt phase transitions in community detectability for spectral community detection under a Erdos-Renyi type random noise model. Our network model includes the widely used stochastic block model \cite{Holland83} as a special case. We show that
at some critical value of random external edge connection probability the community detection performance transitions from almost perfect detectability to low detectability in the large network limit (large $n$). We provide asymptotic upper and lower bounds on this critical value. The bounds become equal to each other when these two community sizes are identical.
This framework can be generalized to community detection on more than two communities by aggregating multiple communities into two larger communities.

We use simulated networks to validate the asymptotic expressions for the phase transitions.
Using our theory, we propose an empirical estimator of the critical phase transition threshold that can be applied to data. These empirical estimates are used to test whether the detector is operating in a reliable detection regime, i.e., below the phase transition threshold.

\section{Network Model and Related Works}

Consider two arbitrarily connected communities with internal adjacency matrices $\ASone$ and $\AStwo$ and network sizes $\none$ and $\ntwo$, respectively. The external connections between these two communities are characterized by an $\none \times \ntwo$ adjacency matrix $\bCS$,
where each entry in $\bCS$ is a Bernoulli($p$) random variable.
Let $n=\none+\ntwo$. The overall $n \times n$ adjacency matrix of the community structure can be represented as
\begin{align}
\label{eqn_asym_block_model}
\bAS = \begin{bmatrix}
       \ASone & \bCS           \\
       \bCS^T           & \AStwo
     \end{bmatrix}.
\end{align}
The widely used stochastic block model \cite{Holland83} is a special case of (\ref{eqn_asym_block_model}) when the two community structures are generated by connected Erdos-Renyi random graphs parameterized by the within-community connection probability $p_i$ ($i=1,2$).
Our network model is more general since we only assume random connection probability $p$ on the external edges and we allow the within-community adjacency matrices $\bAS_i$ to be arbitrary. In this paper we consider the noisy setting in which the adjacency matrix $\bAS$ is corrupted by a random adjacency matrix $\bAN$ such that the observed adjacency matrix is $\bA=\bAS+\bAN$. The adjacency matrix $\bAN$ is generated by a Erdos-Renyi random graph with edge connection probability $q$. Note that this model only allows random insertions and not deletions of edges.

Community detectability has been studied under the stochastic block model with restricted assumptions such as $\none=\ntwo$, $\pone=\ptwo$ and fixed average degree as the network size $n$ increases \cite{Decelle11,Nadakuditi12Detecability,Krzakala13,Radicchi13_hetero,Radicchi14}.
The planted clique detection problem in \cite{Nadakuditi12plant} is a special case of the stochastic block model when $p_1=1$ and $p_2=p$.
A less restricted stochastic block model is studied in \cite{CPY14modularity} where a universal phase transition in community detectability is established for which the critical value does not depend on the community sizes.
A similar model to our network model is studied in \cite{Radicchi13} for interconnected networks. However, in \cite{Radicchi13} the subnetworks are of equal size and the external edges are known (i.e., non-random). Phase transitions in spectral community detection under noiseless network setting is studied in \cite{CPY14spectral}.

\section{Phase Transition Analysis}
\label{sec_system}
Let $\bone_{n_i}$ be the $n_i$-dimensional all-one vector and let $\DSone=\text{diag}\left(\bCS\onentwo\right)$ and $\DStwo=\text{diag}\left(\bCS^T\onenone\right)$.
The graph Laplacian matrix of the noiseless graph can be represented as
\begin{align}
\label{eqn_Laplacian_block}
\bLS = \begin{bmatrix}
       \LSone+\DSone & -\bCS           \\
       -\bCS^T           & \LStwo+\DStwo
     \end{bmatrix},
\end{align}
where $\bLS_i$ is the graph Laplacian matrix of $i$-th community.
Similarly, the graph Laplacian matrix of the noise matrix can be represented as
\begin{align}
\label{eqn_Laplacian_block_noise}
\bLN = \begin{bmatrix}
       \LNone+\DNone & -\bCN           \\
       -\bCN^T           & \LNtwo+\DNtwo
     \end{bmatrix},
\end{align}
where $\bLN_i$ is the graph Laplacian matrix of the noise matrix in $i$-th community, $\bCN$ is the adjacency matrix of noisy edges between two communities,  $\DNone=\text{diag}\left(\bCN\onentwo\right)$ and $\DNtwo=\text{diag}\left(\bCN^T\onenone\right)$.
Therefore the overall graph Laplacian matrix is $\bL=\bLS+\bLN$.

Let $\mathbf{x}=[\xone~\xtwo]^T$, where $\xone \in \mathbb{R}^\none$ and $\xtwo \in \mathbb{R}^\ntwo$. By (\ref{eqn_alge}) we have
$\lambda_2(\mathbf{L})=\min_{\bx} \bx^T \mathbf{L} \bx$ subject to the constraints $\xone^T\xone+\xtwo^T\xtwo=1$ and $\xone^T\onenone+\xtwo^T\onentwo=0$. Using Lagrange multipliers $\mu$, $\nu$ and (\ref{eqn_Laplacian_block}), the Fiedler vector $\by=[\yone~\ytwo]^T$ of $\mathbf{L}$, with $\yone \in \mathbb{R}^\none$ and $\yone \in \mathbb{R}^\ntwo$, satisfies $\by = \arg \min_{\bx} \Gamma(\bx)$, where
\begin{align}
\label{eqn_Lagrange}
\Gamma(\bx)&=\xone^T (\LSone+\DSone+\LNone+\DNone) \xone-2\xone^T (\bCS+\bCN) \xtwo  \nonumber \\
&~~~+ \xtwo^T (\LStwo+\DStwo+\LNtwo+\DNtwo) \xtwo \nonumber \\
&~~~-\mu(\xone^T\xone+\xtwo^T\xtwo-1)-\nu (\xone^T\onenone+\xtwo^T\onentwo).
\end{align}
Differentiating (\ref{eqn_Lagrange}) with respect to $\xone$ and $\xtwo$ respectively, and substituting $\by$ to the equations, we obtain
\begin{align}
\label{eqn_Lagrange1}
&2(\LSone+\DSone+\LNone+\DNone) \yone -2(\bCS+\bCN) \ytwo - 2\mu \yone -\nu \onenone \nonumber\\
&=\mathbf{0}_{\none}, \\
\label{eqn_Lagrange2}
&2(\LStwo+\DStwo+\LNtwo+\DNtwo) \ytwo -2(\bCS+\bCN)^T \yone - 2\mu \ytwo -\nu \onentwo \nonumber\\
&=\mathbf{0}_{\ntwo}.
\end{align}
Left multiplying (\ref{eqn_Lagrange1}) by $\onenone^T$ and left multiplying (\ref{eqn_Lagrange2}) by $\onentwo^T$, we have
\begin{align}
\label{eqn_Lagrange3}
&2\onenone^T(\DSone+\DNone) \yone -2\onenone^T(\bCS+\bCN) \ytwo - 2\mu \onenone^T\yone -\nu \none \nonumber\\
&=0, \\
\label{eqn_Lagrange4}
&2\onentwo^T(\DStwo+\DNtwo) \ytwo -2\onentwo^T(\bCS+\bCN)^T \yone - 2\mu \onentwo^T\ytwo -\nu \ntwo \nonumber\\
&=0.
\end{align}
Since by definition $\onenone^T\DSone=\onentwo^T\bCS^T$, $\onenone^T\bCS=\onentwo^T\DStwo$, $\onenone^T\DNone=\onentwo^T\bCN^T$ and $\onenone^T\bCN=\onentwo^T\DNtwo$, adding (\ref{eqn_Lagrange3}) and (\ref{eqn_Lagrange4}) we obtain
$\nu=-\frac{2\mu}{n} (\yone^T\onenone+\ytwo^T\onentwo)=0$ by the fact that the Fiedler vector $\by$ has the property $\by^T \bone=0$.
Applying $\nu=0$ and left multiplying  (\ref{eqn_Lagrange1}) by $\yone^T$ and left multiplying (\ref{eqn_Lagrange2}) by $\ytwo^T$, we have
\begin{align}
\label{eqn_Lagrange5}
&\yone^T (\LSone+\DSone+\LNone+\DNone) \yone -\yone^T (\bCS+\bCN) \ytwo-\mu \yone^T \yone \nonumber\\
&=0,\\
\label{eqn_Lagrange6}
&\ytwo^T (\LStwo+\DStwo+\LNtwo+\DNtwo) \ytwo -\ytwo^T (\bCS+\bCN)^T \yone-\mu \ytwo^T \ytwo \nonumber\\
&=0.
\end{align}
Adding (\ref{eqn_Lagrange5}) and (\ref{eqn_Lagrange6}) and by (\ref{eqn_alge}) and (\ref{eqn_Laplacian_block}) we obtain $\mu=\lambda_2(\mathbf{L})$.

Let $\bCSbar=p \onenone \onentwo^T$, a matrix whose elements are the means of entries in $\bCS$. Let $\sigma_i(\mathbf{M})$ denote the $i$-th largest singular value of a rectangular matrix $\mathbf{M}$\footnote{Note that for convenience, we use $\lambda_i({\mathbf{M}}_1)$ to denote the $i$-th smallest eigenvalue of a square matrix $\mathbf{M}_1$ and use $\sigma_i({\mathbf{M}}_2)$ to denote the $i$-th largest singular value of a rectangular matrix $\mathbf{M}_2$.}
 and write
$\bCS=\bCSbar+\bDeltaS$,
where $\bDeltaS=\bCS-\bCSbar$. By Latala's theorem \cite{Latala05},
$\mathbb{E} \Lb \sigma_1\lb \frac{\bDeltaS}{\sqrt{\none \ntwo}} \rb \Rb \rightarrow 0$.
This is proved in Appendix VII-A of \cite{CPY14spectral}.
Furthermore, by Talagrand's concentration inequality \cite{Talagrand95}, almost surely,
\begin{align}
\label{eqn_Talagrand}
 \sigma_1\lb \frac{\bCS}{\sqrt{\none \ntwo}} \rb \ra p;~
 \sigma_i\lb \frac{\bCS}{\sqrt{\none \ntwo}} \rb \ra 0~~\forall~i \geq 2
\end{align}
when $\none,\ntwo \rightarrow \infty$ and $\frac{\none}{\ntwo} \rightarrow c>0$. This is proved in Appendix VII-B of \cite{CPY14spectral}. Note that the convergence rate is maximal when $\none=\ntwo$ because $\none+\ntwo \geq 2\sqrt{\none \ntwo}$ and the equality holds if $\none=\ntwo$. Similarly, let $\bCNbar=q \onenone \onentwo^T$, a matrix whose elements are the means of entries in $\bAN$. We have
 $\sigma_1\lb \frac{\bCN}{\sqrt{\none \ntwo}} \rb \ra q$ and
 $\sigma_i\lb \frac{\bCN}{\sqrt{\none \ntwo}} \rb \ra 0~~\forall~i \geq 2$
when $\none,\ntwo \rightarrow \infty$ and $\frac{\none}{\ntwo} \rightarrow c>0$.

As proved in \cite{BenaychGeorges12}, the singular vectors of $\bCS$ ($\bCN$) and $\bCSbar$ ($\bCNbar$) are close to each other in the sense that the squared inner product of their left/right singular vectors converges to $1$ almost surely when $\sqrt{\none \ntwo} p \rightarrow \infty$ ($\sqrt{\none \ntwo} q \rightarrow \infty$).
Consequently, we have, almost surely,
\begin{align}
\label{eqn_D1}
&\frac{(\DSone+\DNone) \onenone}{\ntwo}=\frac{(\bCS+\bCN) \onentwo}{\ntwo} \ra (p+q) \onenone; \\
\label{eqn_D2}
&\frac{(\DStwo+\DNtwo) \onentwo}{\none}=\frac{(\bCS+\bCN)^T\onenone}{\none}  \ra (p+q) \onentwo.
\end{align}

Applying (\ref{eqn_Talagrand}), (\ref{eqn_D1}) and (\ref{eqn_D2}) to (\ref{eqn_Lagrange3}) and (\ref{eqn_Lagrange4}) and recalling that $\nu=0$ and $\frac{\none}{\ntwo}=c>0$,
 we have, almost surely,
\begin{align}
\label{eqn_Lagrange7}
&\frac{1}{\sqrt{c}}(p+q) \onenone^T \yone-\sqrt{c} (p+q) \onentwo^T \ytwo - \frac{\mu \onenone^T \yone}{\sqrt{\none\ntwo}} \ra 0;  \\
\label{eqn_Lagrange8}
&\sqrt{c} (p+q) \onentwo^T \ytwo-\frac{1}{\sqrt{c}} (p+q) \onenone^T \yone - \frac{\mu \onentwo^T \ytwo}{\sqrt{\none\ntwo}} \ra 0.
\end{align}
By the fact that $\onenone^T \yone+\onentwo^T \ytwo=0$, we have, almost surely,
\begin{align}
\label{eqn_Lagrange9}
&\lb \sqrt{c} + \frac{1}{\sqrt{c}} \rb \left(p+q-\frac{\mu}{n} \right) \onenone^T \yone \ra 0; \\
&\lb \sqrt{c} + \frac{1}{\sqrt{c}} \rb \left(p+q-\frac{\mu}{n} \right) \onentwo^T \ytwo \ra 0.
\end{align}
Consequently, as $\mu=\lambda_2(\bL)$, at least one of the two cases have to be satisfied:
\begin{align}
\label{eqn_Lagrange11}
& \text{Case 1:}~\frac{\lambda_2(\mathbf{L})}{n}\overset{\text{a.s.}}{\longrightarrow}p+q=:t, \\
\label{eqn_Lagrange12}
& \text{Case 2:}~\onenone^T \yone \ra 0~~\text{and}~~\onentwo^T \ytwo \ra 0~~\text{almost~surely}.
\end{align}

We will show that the algebraic connectivity $\lambda_2(\bL)/n$ and the Fiedler vector $\by$ undergo a phase transition between Case 1 and Case 2 as a function of $t=p+q$. That is, a transition from Case 1 to Case 2 occurs when $p$ exceeds a certain threshold $p^*$.
In Case 1, observe that asymptotically $\frac{\lambda_2(\mathbf{L})}{n}$ grows linearly with $t$ while the asymptotic Fiedler vector remains the same (unique up to its sign). Furthermore, from (\ref{eqn_Lagrange5}), (\ref{eqn_Lagrange6}), (\ref{eqn_Talagrand}), (\ref{eqn_Lagrange11}), $\mu=\lambda_2(\bL)$ and $\onenone^T \yone+\onentwo^T \ytwo=0$, the Fielder vector $\by$ in Case 1 has the following property. Almost surely,
\begin{align}
\label{eqn_Lagrange13}
&\frac{\yone^T (\LSone+\LNone) \yone}{\sqrt{\none \ntwo}} + \frac{p+q}{\sqrt{\none \ntwo}}(\onenone^T \yone)^2-\sqrt{c} (p+q)\yone^T\yone \ra 0, \\
\label{eqn_Lagrange14}
&\frac{\ytwo^T (\LStwo+\LNtwo) \ytwo}{\sqrt{\none \ntwo}} + \frac{p+q}{\sqrt{\none \ntwo}}(\onenone^T \yone)^2-\frac{1}{\sqrt{c}} (p+q) \ytwo^T\ytwo \ra 0.
\end{align}
Adding (\ref{eqn_Lagrange13}) and (\ref{eqn_Lagrange14}), we have
\begin{align}
\label{eqn_Lagrange15}
&\frac{1}{\sqrt{\none \ntwo}}\left[\yone^T (\LSone+\LNone) \yone + \ytwo^T (\LStwo+\LNtwo) \ytwo \right] + \nonumber \\
&\Lb \frac{2(\onenone^T \yone)^2}{\sqrt{\none \ntwo}}-\left(\sqrt{c} \yone^T\yone+\frac{1}{\sqrt{c}}\ytwo^T\ytwo\right) \Rb (p+q)
\overset{\text{a.s.}}{\longrightarrow}0.
\end{align}
As the two bracketed terms in (\ref{eqn_Lagrange15}) converge to finite constants for all $t=p+q$ in Case 1; almost surely,
\begin{align}
\label{eqn_Lagrange16}
&\frac{1}{\sqrt{\none \ntwo}}\left[\yone^T (\LSone+\LNone) \yone + \ytwo^T (\LStwo+\LNtwo) \ytwo \right] \ra 0; \\
&\frac{2(\onenone^T \yone)^2}{\sqrt{\none \ntwo}}-\left(\sqrt{c} \yone^T\yone+\frac{1}{\sqrt{c}}\ytwo^T\ytwo\right) \ra 0.
\end{align}
By the PSD property of the graph Laplacian matrix, $\yone^T (\LSone+\LNone) \yone + \ytwo^T (\LStwo+\LNtwo) \ytwo>0$ if and only if $\yone$ and $\ytwo$ are not constant vectors. Therefore (\ref{eqn_Lagrange16}) implies $\yone$ and $\ytwo$ converge to constant vectors. By the constraints $\yone^T\yone+\ytwo^T\ytwo=1$ and $\onenone^T\yone+\onentwo^T\ytwo=0$, we have, almost surely,
\begin{align}
\label{eqn_Lagrange17}
\sqrt{\frac{n \none}{\ntwo}} \yone \ra \pm \onenone
~~\text{and}~~
\sqrt{\frac{n \ntwo}{\none}} \ytwo \ra \mp \onentwo.
\end{align}
Consequently, in Case 1 $\yone$ and $\ytwo$ tend to be constant vectors with opposite signs.
More importantly,  (\ref{eqn_Lagrange17}) suggests a phase transition in spectral community detectability.
In Case 1,
spectral clustering can almost correctly identify these two communities since $\yone$ and $\ytwo$ are constant vectors with opposite signs. On the other hand, in Case 2, $\onenone^T \yone \ra 0$ and $\onentwo^T \ytwo \ra 0$ almost surely. The entries of $\yone$ and $\ytwo$
tend to have opposite signs in their entries. Therefore in Case 2 spectral clustering results in very poor community detection.

\begin{table*}[t]
\renewcommand{\arraystretch}{1.3}
    \begin{minipage}[b]{.45\textwidth}
        \includegraphics[width=3.2in]{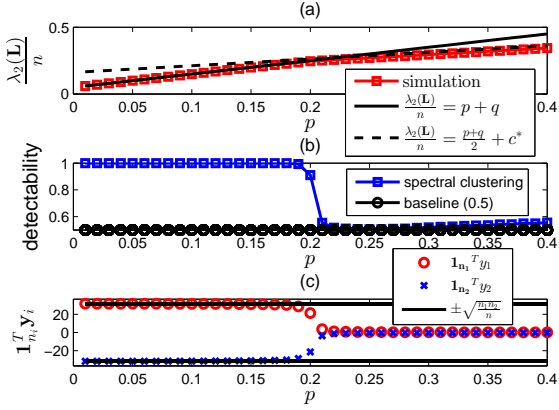}
    \captionof{figure}{Two communities generated by the stochastic block model \cite{Holland83}. The results are averaged over 100 trials. $n_1=n_2=2000$, $p_1=p_2=0.25$, and $q=0.05$. The theoretical critical value from (\ref{eqn_critical_value_equal_size}) is $p^*=0.2229$.}
    \label{Fig_Spec_2000_2000_500_500_q_005}
        \par\vspace{0pt}
    \end{minipage}
    \begin{minipage}[b]{.55\textwidth }
    \centering
        \begin {footnotesize}
\begin{tabular}{|l|l|l|l|l|l|l|}
\hline
\multicolumn{2}{|l|}{~~~~~~~~~~~noise level ($q$)}  & 0                         & 0.002                          & 0.01                           & 0.05                           & 0.1                            \\ \hline
                                       & mean       & 0.8571                    & 0.8548                         & 0.8004                         & 0.6325                         & 0.5038                         \\ \cline{2-7}
\multirow{-2}{*}{detectability}        & std        & \cellcolor[HTML]{EFEFEF}0 & \cellcolor[HTML]{EFEFEF}0.006  & \cellcolor[HTML]{EFEFEF}0.1227 & \cellcolor[HTML]{EFEFEF}0.1597 & \cellcolor[HTML]{EFEFEF}0.0823 \\ \hline
                                       & mean       & 0.0127                    & 0.0116                         & 0.0076                         & 0.00016                        & 0                              \\ \cline{2-7}
\multirow{-2}{*}{$\hpLB$}              & std        & \cellcolor[HTML]{EFEFEF}0 & \cellcolor[HTML]{EFEFEF}0.0021 & \cellcolor[HTML]{EFEFEF}0.0039 & \cellcolor[HTML]{EFEFEF}0.001  & \cellcolor[HTML]{EFEFEF}0      \\ \hline
                                       & mean      & 0.0073                    & 0.0095                         & 0.0173                         & 0.0513                         & 0.0835                         \\ \cline{2-7}
\multirow{-2}{*}{$\hp$}                & std        & \cellcolor[HTML]{EFEFEF}0 & \cellcolor[HTML]{EFEFEF}0.001  & \cellcolor[HTML]{EFEFEF}0.0025 & \cellcolor[HTML]{EFEFEF}0.011  & \cellcolor[HTML]{EFEFEF}0.0209 \\ \hline
                                       & mean       & 0.013                     & 0.0124                         & 0.0633                         & 0.1422                         & 0.1494                         \\ \cline{2-7}
\multirow{-2}{*}{$\hpUB$}              & std        & \cellcolor[HTML]{EFEFEF}0 & \cellcolor[HTML]{EFEFEF}0.0021 & \cellcolor[HTML]{EFEFEF}0.1493 & \cellcolor[HTML]{EFEFEF}0.3199 & \cellcolor[HTML]{EFEFEF}0.3213 \\ \hline
\multicolumn{2}{|l|}{fraction of $\hp \leq \hpLB$}      & 1                         & 0.98                           & 0.01                           & 0                              & 0                              \\ \hline
\multicolumn{2}{|l|}{fraction of $\hpUB<\hp<\hpUB$} & 0                         & 0.02                           & 0.75                           & 0.2                            & 0.2                            \\ \hline
\multicolumn{2}{|l|}{fraction of $\hp \geq \hpUB$}      & 0                         & 0                              & 0.24                           & 0.8                            & 0.8                            \\ \hline
\end{tabular}
\label{Table_political_book_noisy}
\captionof{table}{\textnormal{Sensitivity of spectral community detection to noisy edge insertions for Amazon American political books co-purchasement data \cite{Newman06PNAS}. The network contains 105 nodes and 441 edges. The oracle detectability is 0.8762. The noisy edges are randomly generated for 100 trials.}}
    \par\vspace{0pt}
    \end {footnotesize}
    \end{minipage}%
\end{table*}

\section{Upper and Lower Bounds on the Critical Value}
\label{sec_bounds}
Next we derive an upper bound on the critical value $p^*$ of the phase transition. From (\ref{eqn_alge}) and (\ref{eqn_Laplacian_block}) we know that
\begin{align}
\label{eqn_alge_express}
\lambda_2(\mathbf{L})&=\yone^T (\LSone+\DSone+\LNone+\DNone) \yone-2\yone^T (\bCS+\bCN) \ytwo  \nonumber \\
 &~~~+ \ytwo^T (\LStwo+\DStwo+\LNtwo+\DNtwo) \ytwo
\end{align}
subject to $\onenone^T\yone+\onentwo^T\ytwo=0$ and $\yone^T\yone+\ytwo^T\ytwo=1$.
In Case 2, since $\onenone^T \yone \ra 0$ and $\onentwo^T \ytwo \ra 0$ almost surely, recalling the definition $\bDeltaS=\bCS-\bCSbar$ and let $\bDeltaN=\bCN-\bCNbar$,
\begin{align}
&\frac{\yone^T (\bCS+\bCN) \ytwo}{\sqrt{\none \ntwo}}
=\frac{\yone^T  (\bCSbar+\bCNbar)\ytwo + \yone^T \bDeltaS \ytwo + \yone^T \bDeltaN \ytwo}{\sqrt{\none \ntwo}}   \nonumber \\
& \leq \frac{\yone^T  (\bCSbar+\bCNbar)\ytwo + \| \yone \|_2 \| \ytwo \|_2 \cdot \Lb\sigma_1(\bDeltaS)+\sigma_1(\bDeltaN) \Rb}{\sqrt{\none \ntwo}} \nonumber\\
&\asconv 0
\end{align}
by the fact that $\sigma_1\lb \frac{\bDeltaS}{\sqrt{\none \ntwo}} \rb \asconv 0$ and $\sigma_1\lb \frac{\bDeltaN}{\sqrt{\none \ntwo}} \rb \asconv 0$ in Appendix VII-B of \cite{CPY14spectral} and $\bCSbar=p \onenone \onentwo^T$ and $\bCNbar=q \onenone \onentwo^T$. Furthermore, since $\DSone=\text{diag}\left(\bCS\onentwo\right)$, $\DStwo=\text{diag}\left(\bCS^T\onenone\right)$, $\DNone=\text{diag}\left(\bCN\onentwo\right)$ and $\DNtwo=\text{diag}\left(\bCN^T\onenone\right)$, (\ref{eqn_Talagrand}) gives, almost surely,
\begin{align}
&\frac{1}{\ntwo}\yone^T (\DSone+\DNone) \yone \ra  (p+q) \yone^T \yone; \\
&\frac{1}{\none}\yone^T (\DStwo+\DNtwo) \yone \ra  (p+q) \ytwo^T \ytwo.
\end{align}
Therefore in Case 2 we have
\begin{align}
\label{eqn_case2_alge}
\frac{\lambda_2(\mathbf{L})}{n} \asconv \min_{\bx \in \mathcal{S}} \LB \frac{\xone^T \Lone \xone + \xtwo^T \Ltwo \xtwo + \ntwo t \xone^T \xone + \none t \xtwo^T \xtwo}{n}  \RB,
\end{align}
where $\bL_i=\bLS_i+\bLN_i$, $t=p+q$, and
\begin{align}
&\mathcal{S}=\LB \bx=[\xone~\xtwo]^T:\onenone^T\xone=\onentwo^T\xtwo=0,~\xone^T\xone+\xtwo^T\xtwo=1 \RB.
\end{align}
Define two sets
\begin{align}
&\mathcal{S}_1=\LB \bx:\onenone^T\xone=\onentwo^T\xtwo=0,~\xone^T\xone=1,~\xtwo^T\xtwo=0 \RB;  \\
&\mathcal{S}_2=\LB \bx:\onenone^T\xone=\onentwo^T\xtwo=0,~\xone^T\xone=0,~\xtwo^T\xtwo=1 \RB,
\end{align}
and define
\begin{align}
\mu_i(\bL)= \min_{\bx \in \mathcal{S}_i} \LB \frac{\xone^T \Lone \xone + \xtwo^T \Ltwo \xtwo + \ntwo t \xone^T \xone + \none t \xtwo^T \xtwo}{n} \RB.
\end{align}
Since $\mathcal{S}_1,\mathcal{S}_2 \subseteq \mathcal{S}$, we have, almost surely,
\begin{align}
&\frac{\lambda_2(\mathbf{L})}{n} \leq \min \LB  \mu_1(\bL) , \mu_2(\bL)\RB \nonumber \\
& = \min \LB  \frac{\lambda_2(\bL_1) + \ntwo t}{n} , \frac{\lambda_2(\bL_2)+\none t}{n}\RB \nonumber \\
& = \frac{t}{2}+\frac{\lambda_2(\bL_1) +\lambda_2(\bL_2)-\left| \lambda_2(\bL_1) -\lambda_2(\bL_2)+(\ntwo-\none) t \right|}{2n} \nonumber \\
& \leq \frac{t}{2}+ \frac{|\none-\ntwo|t}{2n} + \frac{\lambda_2(\bL_1) +\lambda_2(\bL_2)-\left| \lambda_2(\bL_1) -\lambda_2(\bL_2) \right|}{2n},
\label{eqn_critical_value_UB}
\end{align}
where we use the facts that $\min\{a,b\}=\frac{a+b-|a-b|}{2}$ and $|a-b| \geq |a|-|b|$. Note that the last equality in (\ref{eqn_critical_value_UB}) holds if $\none=\ntwo$.  Let $t^*=p^*+q$ be the critical value for phase transition from Case 1 to Case 2. There is a phase transition on the asymptotic value of $\frac{\lambda_2(\bL)}{n}$ since the slope of $\frac{\lambda_2(\bL)}{n}$ converges to 1 almost surely when $t \leq t^*$, whereas from (\ref{eqn_critical_value_UB}) $\frac{\lambda_2(\bL)}{n}-t \leq \frac{\lb |\none-\ntwo|-n \rb t}{2n} + \frac{\lambda_2(\bL_1) +\lambda_2(\bL_2)-\left| \lambda_2(\bL_1) -\lambda_2(\bL_2) \right|}{2n}$ when $t \geq t^*$.
From (\ref{eqn_Lagrange11}), we obtain an asymptotic upper bound $\pUB$ on the critical value $p^*$ by substituting $t^*=p^*+q$ to (\ref{eqn_critical_value_UB}).
\begin{align}
\label{p_UB}
\pUB  = \frac{\lambda_2(\bL_1) +\lambda_2(\bL_2)-\left| \lambda_2(\bL_1) -\lambda_2(\bL_2) \right|}{n-|\none-\ntwo|}-q.
\end{align}

To derive a lower bound on $p^*$, we have that in Case 2,
\begin{align}
&\frac{\lambda_2(\bL)}{n} \asconv \min_{\bx \in \mathcal{S}} \LB \frac{\xone^T \Lone \xone + \xtwo^T \Ltwo \xtwo + \ntwo p \xone^T \xone + \none p \xtwo^T \xtwo}{n}   \RB
\nonumber \\
\label{eqn_critical_value_LB_2}
& \geq \min_{\bx \in {S}} \LB \frac{\xone^T \Lone \xone + \xtwo^T \Ltwo \xtwo}{n}  \RB+ \min_{\bx \in S} \LB \frac{\ntwo t \xone^T \xone + \none t \xtwo^T \xtwo}{n}  \RB
\\
&= \min\LB \frac{\lambda_2(\bL_1)}{n},\frac{\lambda_2(\bL_2)}{n} \RB + \min \LB \frac{\none t}{n}, \frac{\ntwo t}{n}\RB. \nonumber \\
\label{eqn_critical_value_LB}
&= \frac{t}{2}-\frac{|\none-\ntwo| t}{2n}+ \frac{\lambda_2(\bL_1) +\lambda_2(\bL_2)-\left| \lambda_2(\bL_1) -\lambda_2(\bL_2) \right|}{2n}.
\end{align}
Substituting $t^*=p^*+q$ to (\ref{eqn_critical_value_LB}), we obtain an asymptotic lower bound $\pLB$ on the critical value $p^*$.
\begin{align}
\label{p_LB}
\pLB = \frac{\lambda_2(\bL_1) +\lambda_2(\bL_2)-\left| \lambda_2(\bL_1) -\lambda_2(\bL_2) \right|}{n+|\none-\ntwo|}-q.
\end{align}\underline{}
Note that when $\none=\ntwo$, the equality in (\ref{eqn_critical_value_LB_2}) holds. This means when $\none=\ntwo$, $\frac{\lambda_2(\bL)}{n} \asconv \frac{t}{2} +\frac{\lambda_2(\bL_1) +\lambda_2(\bL_2)-\left| \lambda_2(\bL_1) -\lambda_2(\bL_2) \right|}{2n}=:\frac{t}{2} + c^*$ in Case 2,
and the critical value
\begin{align}
\label{eqn_critical_value_equal_size}
p^* \asconv \frac{\lambda_2(\bL_1) +\lambda_2(\bL_2)-\left| \lambda_2(\bL_1) -\lambda_2(\bL_2) \right|}{n}-q.
\end{align}

Here we derive the bounds on the critical value $p^*$ for the stochastic block model, where the internal adjacency matrix $\bA_i$ in (\ref{eqn_asym_block_model}) is generated by a Erdos-Renyi random graph with edge connection probability $p_i$. It is proved in Appendix VII-C of \cite{CPY14spectral} that
$\lambda_2 \lb \frac{\bL_i}{n_i} \rb \asconv p_i+q$.  Therefore $\pUB=\frac{c \pone + \ptwo-|c \pone-\ptwo+(c-1)q|-|1-c|q}{1+c-|1-c|}$ and $\pLB=\frac{c \pone + \ptwo-|c \pone-\ptwo+(c-1)q|-|1-c|q}{1+c+|1-c|}$. When $\none = \ntwo$ (i.e., $c=1$), the critical value
   $p^* \asconv \frac{\pone + \ptwo -|\pone - \ptwo|}{2}$. This suggests that in the largest network limit when $n \ra \infty$ and $c=1$ the performance of spectral community detection is independent of the noise parameter $q$.

\section{Performance Evaluation}
\label{Sec_performance}
\subsection{Simulated Networks}
We use the stochastic block model \cite{Holland83} to generate network graphs for community detection.
The detectability is defined as the fraction of nodes that are correctly identified and the baseline detectability is 0.5 for random guesses.
In Fig. 1, when $\pone=\ptwo=0.25$, $\none=\ntwo=2000$ and $q=0.05$, the theoretical critical value from (\ref{eqn_critical_value_equal_size}) is $p^*=0.2229$. Note that $p^*$ will converge to $0.25$ as we increase $n$ as predicted in Sec. \ref{sec_bounds}.

Fig. \ref{Fig_Spec_2000_2000_500_500_q_005} (a) verifies the phase transition in $\frac{\lambda_2(\bL)}{n}$ empirically confirming that $\frac{\lambda_2(\bL)}{n}$ approaches $p+q$ when $p \leq p^*$ and  $\frac{\lambda_2(\bL)}{n}$ approaches $\frac{p+q}{2}+c^*$ when $p > p^*$, where $c^*=\frac{\lambda_2(\bL_1) +\lambda_2(\bL_2)-\left| \lambda_2(\bL_1) -\lambda_2(\bL_2) \right|}{2n}$.  Fig. \ref{Fig_Spec_2000_2000_500_500_q_005} (b) shows that the community detectability transitions from almost perfect detectability when $p<p^*$ to low detectability when $p > p^*$. Moreover, as derived in (\ref{eqn_Lagrange17}), the Fiedler vector components $\yone$ and $\ytwo$ are constant vectors with opposite signs for $p < p^*$, and $\onenone^T \yone \ra 0$ and $\onentwo^T \ytwo \ra 0$ for $p > p^*$, as shown in Fig. \ref{Fig_Spec_2000_2000_500_500_q_005} (c).

\subsection{Empirical Estimators of Phase Transition Bounds on Real-world Dataset}
Here we show that the critical phase transition threshold $p^*$ can be empirically estimated to empirically test the reliability of spectral community detection.
Let $\widehat{\mathbf{L}}_i$ be the graph Laplacian matrix of the estimated community $i$ obtained by applying spectral clustering to the observed adjacency matrix $\bA$ and let $\widehat{n}_i$ denote the estimated network size of community $i$. Using (\ref{p_UB}) and (\ref{p_LB}), the empirical estimators of these parameters are defined as
\begin{align}
\hp&=\text{number of identified external edges}/\hnone \hntwo, \\
\hpLB&=\frac{\lambda_2(\hLone) +\lambda_2(\hLtwo)-\left| \lambda_2(\hLone) -\lambda_2(\hLtwo) \right|}{n+|\hnone-\hntwo|}, \\
\hpUB&=\frac{\lambda_2(\hLone) +\lambda_2(\hLtwo)-\left| \lambda_2(\hLone) -\lambda_2(\hLtwo) \right|}{n-|\hnone-\hntwo|}.
\end{align}
Based on these empirical estimates, the performance of community detection can be classified into three categories. If $\hp \leq \hpLB$, the network is in the reliable detection region. If $\hpLB < \hp < \hpUB$, the network is in the intermediate detection region. If $\hp \geq \hpUB$, the network is in the unreliable detection region.

The co-purchasement data between 105 American political books sold on Amazon \cite{Newman06PNAS} are used to estimate the parameters $\pLB$, $\pUB$ and $p$.
For the corresponding network graph nodes represent political books and edges represent co-purchasements.
An edge exists between two books if they are frequently purchased by the same buyer. Three labels, \emph{liberal}, \emph{conservative} and \emph{neutral}, were determined by Newman \cite{Newman06PNAS}. We perform community detection by separating the books into two groups since there are only 13 books with neutral labels (i.e., the oracle detectability is 0.8762). To investigate the sensitivity of spectral community detection to noisy edge insertions, for each edge not present in the original graph, an edge is added with probability $q$. The community detection results are summarized in Table I. Observe that for small $q$ ($q$=0 or 0.002) the network is mostly in the reliable detection region ($\hp < \hpLB$), which indicates that spectral community detection achieves high detectability. When $q=0.01$, the network is mostly in the intermediate detection region ($\hpLB < \hp < \hpUB$), indicating that the community detectability has large variation. When $q$ is large ($q$=0.05 or 0.1), the network is mostly in the unreliable detection region resulting in low detectability. The large standard deviation of $\hpUB$ for large $q$ is due to the fact that spectral community detection may mistakenly detect two communities with extremely imbalanced community sizes such that the denominator of the estimator $\hpUB$ is small.

\section{Conclusion}
We establish asymptotic phase transition bounds on the critical value $p^*$ under a general network setting corrupted by a Erdos-Renyi type noise model. The communities are proven to be almost perfectly detectable below the phase transition threshold and to be undetectable above the phase transition threshold. The phase transition bounds are used to establish empirical estimators to evaluate the reliability of spectral community detection, where the detector is said to be operating in the reliable, intermediate, or unreliable detection regime based on the empirical estimates. Simulated networks generated by the stochastic block model validate the phase transition theory for community detectability. An empirical estimator of the phase transition is proposed that can be used to explore sensitivity of the spectral community detection algorithm on real data.

\clearpage
\bibliographystyle{IEEEtran}
\bibliography{IEEEabrv,network_identify}

\end{document}